\documentclass{article}
\usepackage{graphicx}
\usepackage{authblk}
\emergencystretch=\maxdimen
\usepackage[hidelinks]{hyperref}
\usepackage[utf8]{inputenc}
\usepackage{amsfonts}
\usepackage{float}
\usepackage[mathscr]{eucal}
\usepackage{algorithm}
\usepackage{algpseudocode}
\usepackage{amsmath}
\usepackage{xcolor}
\usepackage{amssymb}
\usepackage[british]{babel}

\newtheorem{definition}{Definition}
\usepackage[margin=1in]{geometry}
\usepackage{draftwatermark}
\SetWatermarkText{DRAFT}
\SetWatermarkLightness{0.9}
\SetWatermarkScale{3}

\title{Intermediating DFMM Asset (IDA)}
\author{Arman Abgaryan, Utkarsh Sharma}
\affil{Supra DeFi Research}

\date{August 2023}

\begin{document}

\maketitle

\begin{abstract}    
    The Dynamic Function Market Maker (DFMM) \cite{abgaryan2023dynamic} introduced a fully automated framework for operating a multi-asset market, wherein an algorithmic accounting asset was used to connect different liquidity pools and ensure efficient rebalancing of risks, and internal accounting processes. In the DFMM design, this asset was not tradaeble; however, in this work, we explore the characteristics of this asset, if it were to be made tradeable. Named the Intermediating DFMM Asset (IDA), this asset serves as a unit of account in cross-chain finance, functioning as an intermediating asset for predictable budgeting, and efficient multichain transfers and settlements. Harnessing its robust liquidity as the key counterpart asset in DFMM, it achieves capital efficiency through the strategic repurposing of its asset base, while simultaneously mitigating risk via the dynamic optimisation of its multicollateral foundation. We outline key characteristics of the proposed asset, unique risk mitigation aspects enabled by the adopting AMM (DFMM), and control levers enabling the protocol's tactical asset and liability management toolkit to harmonise the asset's objectives with its real-world realisation, through a novel prudential market operation to incentivise productive use of a finite asset and dynamic AMM fee to ensure alignment of behaviours. The proposed design has the potential to harmonise the interests of diverse market participants, leading to synergetic reactions to informational flow, aiding IDA protocol in achieving its objectives.
\end{abstract}

\section{Motivating Vision}
    Liquidity in today's digital finance landscape is fragmented, resembling disconnected cities, with distinct routes, making transactions akin to navigating a world of isolated metropolises with scattered maps, yearning for a financial bridge to connect these disjointed hubs, unifying financial highways. To address this, we seek to create an intermediating asset that would bridge the fragmented digital finance, enabling cross-chain and cross-system value connectivity. As a \lq\lq multi-chain bridge asset\rq\rq we envision this asset to be the liquid universal counterpart asset for multi-chain transactions in decentralised finance (DeFi), aggregating liquidity scattered between different chains and protocols,  acting as efficient value routing and settlement mechanism for cross-chain DeFi. We aspire to create an intermediating asset that is an efficient vehicle of value exchange, which is responsibly resilient to evolving market dynamics. Judicious responsiveness to evolving dynamics, risks, and vulnerabilities, without being rigid, promotes sustainability in price discovery. We intend to achieve this by blending a proactive and a rule-based reactionary mechanism with decentralised governance-based interventions, enabling even a fully algorithmic intermediating asset to be managed transparently in exceptional circumstances. We seek this asset to become a high-speed train for cross-chain DeFi and digital financial economy that would connect the scattered cities unifying the pathways of value transfers.

\section{Introduction}\label{sec:intro}
    Dynamic Function Market Maker (DFMM)\cite{abgaryan2023dynamic} utilised an accounting counterpart asset, which enabled liquidity aggregation, by harmonising multi-asset liquidity pools for accounting operations and enabling the protocol to achieve its risk management objectives. In this work, we conceptualise how the counterpart asset can be made tradeable, and delve into its utility, objectives, and risk characteristics. Named Intermediating DFMM Asset (IDA), it uses a tactical asset and liability management policy, comprised of individually dynamic asset and liability management principles. The proposed policy and principles serve the objective of enabling IDA to act as a bridge asset for cross-chain transactions, enabling users to budget and settle transactions, whilst also provisioning the use of backing multi-chain and multi-asset liquidity to support convertibility into a desired alternative.\\
    \\
    We now define the asset that is the focus of this work.

    \begin{definition}[Intermediating DFMM Asset (IDA)]
        The Intermediating DFMM Asset (IDA) is an economic unit of measurement, using its tactical asset \& liability management policy to act as a dependable counterpart asset for internal accounting operations for multi-asset platforms (e.g. an adopting AMM like DFMM), without hindering market forces which facilitate price discovery, considering the denominated asset (or basket of assets) IDA's value is pegged to\footnote{For ease of reference, we will assume IDA to be dynamically pegged to USD.}.\\
        \\
        The asset $IDA(TALM(\mathbf{\mathcal{A}}, \mathbf{\mathcal{L}}))$, where $TALM(.)$ is the tactical asset and liability management policy function, which is a function of a function, comprised of dynamic asset and liability management functions, represented by $\mathbf{\mathcal{A}}$ and $\mathbf{\mathcal{L}}$, respectively.
    \end{definition}

    \noindent
    In essence, the objective of TALM policy implementing the dynamic asset and liability management principles seeks to optimally navigate the three-dimensional objective space, comprised of:
    \begin{enumerate}
        \item \textbf{Decentralisation}, referring to the lack of a centralised issuer, mitigating the risk of a single point of failure.
        \item \textbf{Dependability}, which stems from the fact that IDA is a vehicle of cross-chain value transfer, which is universal, and is intended to be deployed for economically meaningful purposes, and hence, the policies strike a balance between activating its management tools and providing holders with the maximum possible time to deploy them for meaningful economic work, and by choice, convert them to alternative assets\footnote{It is important to note that the goal of dependability is distinct from stability. The system is primarily designed to extend the duration for which users can productively utilise the asset, rather than ensuring or aiming for the long-term stability of the risk-capital used to acquire the intermediating asset.}.
        \item \textbf{Capital Efficiency}, a measure used to quantify the economic utility of every unit of asset supporting IDA.
    \end{enumerate}

    \noindent
    Liabilities on IDA's \lq\lq balance sheet\rq\rq are managed using the dual levers, namely - differential AMM fees (\lq\lq soft\rq\rq lever) and prudential market operations (\lq\lq hard\rq\rq lever). The soft lever influences agent behaviours through the transmission mechanism of individual transactions and trades, whereas hard levers influence behaviours by implying a financial cost by way of converting a portion of assets to a different asset mix, freeing up the finite economic asset for productive use. Note, that the prudential market operations are labelled as a hard lever, because even though IDA's policy would aim to maximise the number of timesteps ($\tau_t$) (stabilising time) that a holder has to convert IDA to another asset in the adopting AMM's (DFMM) pool, which if needed, will forcibly act to take corrective decisions. Complementing these two pillars comprising the dynamic liability management principles, are the two pillars comprising dynamic asset management mechanism, which are market-driven and protocol-driven approaches to rebalancing the asset mix, which provides a mix of proactively action-based and responsive set of rules. These rules use signals gleaned from the secondary liquidity provider market presented in the motivating reference work\cite{abgaryan2023dynamic}, where non-linear risks are priced, through continuous optimisation of the size of sLP margin vaults. Both market-driven and protocol-driven levers facilitate the dynamic optimisation of the asset base that backs IDA, whilst facilitating the strategic reallocation of open inventory positions arising across diverse liquidity pools.\\
    \\
    Although given its economic purpose and core mission, IDA is not a \lq\lq stablecoin\rq\rq, in the literature review that follows, we'll categorise and evaluate designs of existing stablecoins. This is because despite the fundamental differences, stablecoin protocols serve as an analytical benchmark for our work.
      
\section{Literature Review}
    Within Decentralised Finance (DeFi), numerous strategies have emerged to facilitate the price stability of a cryptocurrency known as a \lq\lq stablecoin\rq\rq, maintaining a steady value in relation to its underlying asset, whether it's a commodity or a fiat currency. This endeavour seeks to blend the innate stability of the reference asset, most commonly fiat currencies, with the advantageous features of cryptocurrencies.\\
    \\
    Broadly speaking, these frameworks can be allocated to one of the four different categories, which are as follows:

\begin{enumerate}
    \item \textbf{Collateralisation-based}: Stablecoins issued on the basis to the value of another asset, or basket of assets, fall in the category of collateralisation-based stablecoins, which can broadly be divided into two subcategories:
    \begin{itemize}
        \item \textbf{On-chain}: In this subcategory, we consider stablecoins that are backed by on-chain assets like ETH, ZRX, OMG, etc.\cite{url19,url20} and use those assets to facilitate open market operations which assert the peg. Stablecoins backed by on-chain assets are typically over-collateralised to help mitigate the impact of adverse changes in asset prices, which is similar to the use of special purpose vehicles that over-collateralise their debts to other assets. Despite some adoption, these protocols are rife with a host of risks and vulnerabilities, and anyway, suffer from poor efficiency of capital deployed in overcollateralised vaults.
        \item \textbf{Off-chain}: In this subcategory, we consider stablecoins that are backed by off-chain assets like USD, EUR, gold, etc.\cite{url3, url4, url5, url6}. However, there have been concerns about some projects' ability to maintain reserves (e.g. \cite{ url23}). These types of stablecoin systems are highly centralised, requiring a level of trust in faithful execution of stated policies, and the abilities of those centralised operators to manage existing (and foresee emerging) risks, leaving any stability guarantees offered moot.
    \end{itemize}

    \noindent
    Mechanically, both these categories either require users to purchase the stablecoin, or contribute to the collateral base and then have access to loaned stablecoins, both of which contribute to the asset base required to defend the value of the stablecoin. Whether over-collateralised or not, either on-chain or off-chain, such projects lack in either decentralisation, capital efficiency, or in their ability to stablise their peg to the linked asset. In addressing these challenges alternative models, often termed \lq\lq algorithmic stablecoins\rq\rq, have emerged that do not exclusively rely on external assets to sustain their peg, and can be categorised as follows:
    
    \item \textbf{Rebasing-based}: Stablecoins in this category utilise dynamic supply monetary policies to maintain a price peg by regulating their coin's elastic supply to steer its price towards a predefined level. This category of tokens includes AMPL \cite{kuo2019ampleforth}, which adjusted user balances to influence the price, effectively transferring the coin's volatility from its price to its token supply. These schemes struggle to find a stable equilibrium price for their tokens, and dynamically changing token balances make it difficult for the system to be robustly integrated with other platforms.
    
    \item \textbf{Seigniorage-based}: These stablecoins have protocols (e.g., \cite{cao2018designing, kereiakes2019terra, sams2015note}) that separate it's value's stability and uncertainty, by issuing distinct coins - stable and speculative coin, respectively. Once issued, strategies for their management is particularly noteworthy. For e.g., in the case of one particular project\cite{url25}, when the stablecoin's price fell below the peg, speculative coins were auctioned in exchange for the stablecoin (contracting the supply of the stablecoin), thereby diluting the speculative coin holders. And when the price was above the peg, more stablecoins are introduced into circulation, leading to its inflation. From a speculative token holder perspective, these dynamics can be summarised as the speculative token holders providing price support during inflationary times, and being rewarded, in stable or deflationary conditions. From the protocol's perspective, such systems can be reflective and fragile\cite{briola2023anatomy}, largely due to the fact that they are generally uncollateralised and rely entirely on speculative token holders to act as a lender of last resort.
        
    \item \textbf{Hybrid}: Certain projects have adopted a hybrid approach, amalgamating algorithmic stabilisation techniques with collateral reserves. These hybrid mechanisms strive to diminish the coin's dependence on it's reserve and market sentiment factors. Within this category, various hybrid protocols exhibit traits from more than one stablecoin category\cite{url24, zahnentferner2021djed}. However, it is essential to note that these systems inherit some of the limitations associated with the algorithmic (and collateralisation-based) coins mentioned earlier, e.g. in their capacity to absorb external shocks, such as rapid adoption rate declines and collateral price fluctuations, before potential collapse becomes a concern.
\end{enumerate}

    \noindent
    Simply put, algorithmic stablecoins continue to be seen to be flawed, as one work argues \cite{clements2021built}, because they rely on an impossible trinity of three factors that are hard to control, i.e. (a) demand to support operational stability; (b) independent actors participating in price-stabilising arbitrage; and (c) communication of up-to-date and accurate price, without delay.

\section{Preliminaries}\label{sec:prelims}
    In this section, we recall relevant concepts from the motivating reference work, describing the Dynamic Function Market Maker, which is visually summarised in the schematic below.

    \begin{figure}[H]
        \begin{center}
            \includegraphics[width=5in]{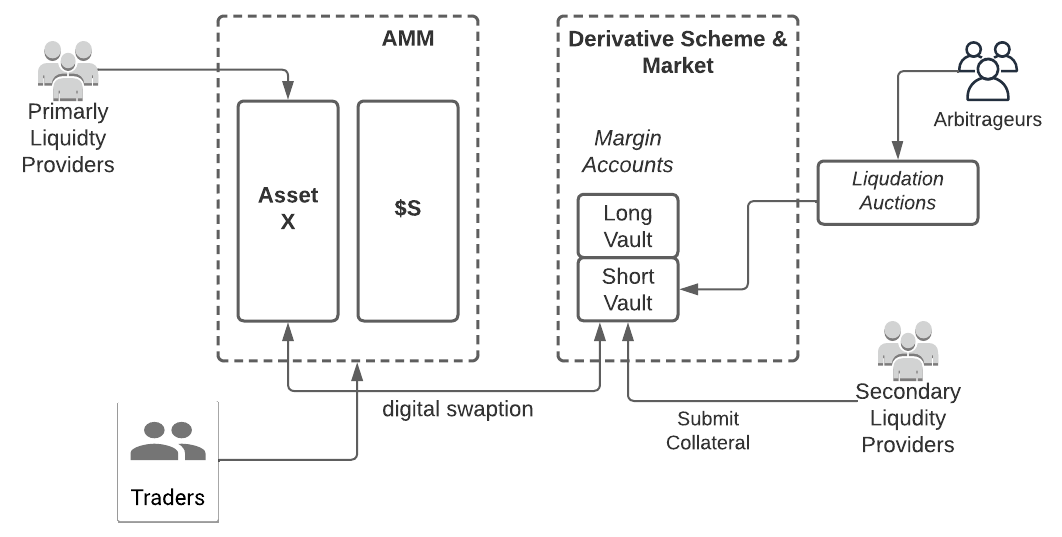}
            \caption{DFMM schematic.}
            \label{dfmmschematics}
        \end{center}
    \end{figure}
    
    \begin{definition}[Digital Swaption]
        A digital swaption is a non-recourse financial instrument, bestowing its holder the right, but not enforcing an obligation, to enter into a total return swap, starting at a pre-specified date in the future, for a fixed maturity and rate, set at the time of inception.
    \end{definition}

    \begin{definition}[Primary Liquidity Providers]
        Primary liquidity providers (pLPs) are market participants who aim to maximise their terminal wealth, denoted as $W_T$ where $T$ represents the terminal time ($T \in (0,\infty)$). It is assumed that pLPs start with an initial wealth value of $W_0$, which is greater than zero. pLPs submit digital assets to an AMM, and by doing so, they provide a stable source of liquidity. Unlike risk-seeking market participants who actively optimise their risk positions, pLPs are primarily passive risk-takers, and hence, in comparison to other types of agents, are the slowest to rebalance their inventory to a new signal. Their focus is on earning an income by providing liquidity, rather than actively seeking risk or maximising returns.
    \end{definition}

    \begin{definition}[Secondary Liquidity Provider (sLP)]
        Secondary liquidity providers (sLP) are market participants similar to pLPs, as defined above. However, unlike pLPs, sLPs do not deposit digital assets into liquidity pools. Instead, they deposit assets as collateral in a specialised margin vault of the adopting AMM's (DFMM) collateral\footnote{Note, that for each asset pool, there are two margin vaults, one for long and one for short.}, offering limited protection to passive pLPs.
    \end{definition}

    \begin{definition}[Rebalancing Premium Auction]
        Rebalancing Premium Auction refers to a Dutch auction in the implementing AMM (DFMM) that is designed to attract stablising flows to the system by providing the smallest possible incentives (rebalancing premium). Its primary objective is to minimise asynchronicity between the local and external markets, while reducing risks and costs to the protocol.
    \end{definition}

    \begin{definition}[Collaterisation Rate]
        The collateralisation rate ($\varrho^i_t$) for a position in asset $i$ at time $t$ represents the percentage of unhedged open inventory that must be held in the vault as collateral to mitigate the inventory risk associated with that position.
    \end{definition}

\section{Tactical Asset \& Liability Management Policy}\label{sec:talm}
    The notion of Asset \& Liability Management (ALM) speaks to the tactical handling of risks stemming from disparities between any institution's assets and liabilities, where a mismanaged mismatch poses a risk to its continuing viability. Similarly, synthetic (digital) asset protocols carry an innate susceptibility to such mismatches, which is compounded by their programmatically static approach to its assessment, management, and mitigation, highlighting the necessity for adaptable strategies to counteract risks and vulnerabilities. This is because these mismatches cannot be entirely negated, as that comes at an unappetising cost to any investor's capital efficiency objectives, therefore, necessitating the need for a dynamic policy to manage assets and liabilities which seek to optimally operate in a multiobjective space.\\
    \\
    Before proceeding to IDA-specific dynamics, let us recall that the adopting AMM (DFMM)\cite{abgaryan2023dynamic} asserted a fundamental condition for its dependability, which sought to assert that LPs can withdraw assets they have deposited, which are a liability for the adopting AMM. This in turn implies the following requirement to be fulfilled:

    \begin{equation}\label{almbalance}
    \begin{split}
     \underbrace{\mathcal{L}^{E^X}_t(\mathcal{I}^X_t) + \mathcal{L}^{E^Y}_t(\mathcal{I}^Y_t)}_\text{assets}   = 
        \overbrace{\mathcal{L}^{E^X}_t(\mathcal{I}^X_{{LP}_t}) + \mathcal{L}^{E^Y}_t(\mathcal{I}^Y_{{LP}_t})}^\text{liabilities},
    \end{split}
    \end{equation}

    \noindent
    where $\mathcal{L}^{E^X}_t(.): V_X \rightarrow V_{S}$ denotes the external liquidity density function for asset $X$ (in IDA), linking asset $X$'s volume to the value of a designated base asset\footnote{The designated base asset is the asset to which IDA is pegged to.}; $\mathcal{I}^X_t$ represents the inventory of assets held in X-th asset's pool; $\mathcal{I}^X_{LP_{t}}$ denotes the inventory owned by liquidity providers in $X$-th asset's pool at time $t$(referred to as primary liquidity providers in the motivating work). Note, that in the reference work, the AMM had in-built mechanisms to help assert its objectives, for e.g., by using collateral deposited in margin vaults of secondary LPs (sLPs). Therefore, DFMM assets can be expressed as follows:
    
    \begin{equation}
          \text{Assets}^{DFMM} =\mathcal{L}^E_t(\mathcal{I}^X_t) + \mathcal{L}^E_t(\mathcal{I}^Y_t)+ \mathcal{L}^{E^X}_t (\mathcal{C}^{{\prime}^X}_t) + \mathcal{L}^{E^Y}_t (\mathcal{C}^{{\prime}^Y}_t)
    \end{equation}
    
    \noindent
    where $\mathcal{C}^{{\prime}^X}_t$ is the collateral submitted to both the short and long vaults for asset $X$.\\
    \\
    Now, for the IDA protocol, its liabilities (which are in addition to the adopting AMM (DFMM's) liabilities) are quantified using the units of IDAs introduced into circulation ($CS^S_t$) \footnote{We denote the IDA asset by $S$}, which have the effect of transforming the adopting AMM's liabilities as follows: 
    \begin{equation*}
        \text{Liabilities}^{DFMM} = \mathcal{L}^{E^X}_t(\mathcal{I}^X_{{LP}_t}) + \mathcal{L}^{E^Y}_t(\mathcal{I}^Y_{{LP}_t}) + CS^S_t
    \end{equation*}
   
\begin{figure}[H]
        \begin{center}
            \includegraphics[width=3in]{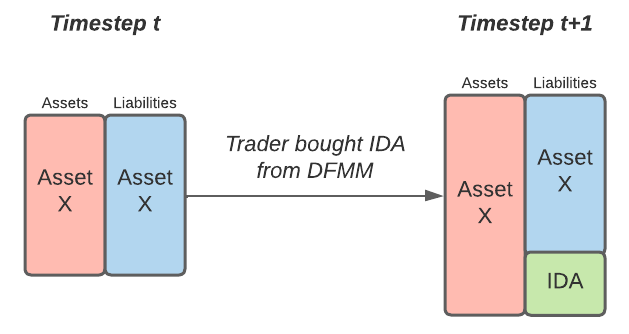}
            \caption{Post IDA accrual balance sheet composition.}
            \label{idabs}
        \end{center}
    \end{figure}

    \noindent
    The introduced circulating supply is affected (increases) by - (i) distribution of AMM fees among its agents; (ii) enhancing treasury with earned fees; and (iii) procurement of IDA from DFMM when the accounting asset is made tradeable\footnote{However note, that if IDA is sold to the adopting AMM (DFMM), IDA's circulating supply decreases.}. For ease of reader's convenience, we illustrate how the system's balance sheet evolves with a purchase of IDA(Fig.\ref{idabs}). Specifically, note, that IDA's TALM seeks to modulate only those liabilities (and resultingly, its assets), which are directly attributable to IDA, as liabilities of the adopting AMM (DFMM) may be larger than IDA's liabilities($\text{Liabilities}^\text{IDA} \subseteq \text{Liabilities}^\text{AMM}$).
    
    \noindent
    Similarly, from the adopting AMM's perspective, the introduction of IDA changes its inventory and asset base as follows:
    
    \begin{equation}
        \text{Assets}^{DFMM} =\underbrace{\mathcal{L}^E_t(\mathcal{I}^X_t) + \mathcal{L}^E_t(\mathcal{I}^Y_t)+ \mathcal{L}^{E^X}_t (\mathcal{C}^{\prime^X}_t) + \mathcal{L}^{E^Y}_t (\mathcal{C}^{\prime^Y}_t)}_\text{before IDA} + \overbrace{ \mathcal{L}^E_t(\Delta \mathcal{I}^X_t) + \mathcal{L}^E_t(\Delta \mathcal{I}^Y_t)}^\text{after IDA},  
    \end{equation} 

    \noindent
    where the $\Delta \mathcal{I}^X_t$ and $\Delta \mathcal{I}^Y_t$ represent the additional inventory submitted to the AMM when IDA is introduced to circulation.\\
    \\
    Now, if Eq.\ref{almbalance} holds, nominal open inventory created due to an increase in IDA's circulating supply can be quantified using the difference between negative and positive open inventory, as follows:
        
    \begin{equation}
        CS = \mathcal{L}^E_t(\Delta \mathcal{I}^X_t) + \mathcal{L}^E_t(\Delta \mathcal{I}^Y_t)= \sum_{i=1}^{n} \mathcal{L}^{E^i}_t (\mathcal{I}^i_t - \mathcal{I}^i_{{LP}_t})- \sum_{j=1}^{m} \mathcal{L}^{E^j}_t (\mathcal{I}^j_{LP_t} - \mathcal{I}^j_{t})
    \end{equation}

    \noindent
    where $i$ denotes the pool which has a positive open inventory position \footnote{Open inventory position refers to the difference between the current inventory of assets available in an AMM's liquidity pool and inventory that is owned by LP in that pool.} and  $j$ denotes the pools which have negative open inventory position; $n$ and $m$ are the number of pools that have positive and negative open inventory position, respectively.\\
    \\
    However, assets that are attributable to IDA will also include a portion of the collateral pool that specifically acts as a buffer to protect the additional liabilities that are created due to the introduction of IDA.  Therefore total assets attributable to  IDA($\text{Assets}^\text{IDA} \subseteq \text{Assets}^\text{AMM}$) can be expressed as follows:
    
    \begin{equation}
        A_t = \sum_{i=1}^{n} \mathcal{L}^{E^i}_t (\mathcal{I}^i_t - \mathcal{I}^i_{{LP}_t})- \sum_{j=1}^{m} \mathcal{L}^{E^j}_t (\mathcal{I}^j_{LP_t} - \mathcal{I}^j_{t}) + \sum^n_{i=1} \mathcal{L}^{E^i}_t (\mathcal{C}^i_t)\cdot \frac{\sum_{i=1}^{n} \mathcal{L}^{E^i}_t (\mathcal{I}^i_t - \mathcal{I}^i_{{LP}_t})- \sum_{j=1}^{m} \mathcal{L}^{E^j}_t (\mathcal{I}^j_{LP_t} - \mathcal{I}^j_{t})}{\sum_{i=1}^{n} \mathcal{L}^{E^i}_t (\mathcal{I}^i_t - \mathcal{I}^i_{{LP}_t})}
    \end{equation}
    
    \noindent
    where $A_t$ represents the nominal value of IDA protocol's assets; and  $\mathcal{C}^i_t$ represents the size of available collateral in the long vault, which is equivalent to $\mathcal{C}^{i}_{{long}_t}$ notation in the reference DFMM work.\\
    \\
    Note, that when assessing the optimality of an asset's balance sheet composition, only assets exclusively designated to safeguard liabilities tied to the issued synthetic asset are considered. Furthermore, although open inventory positions are part of total assets, they are generally deposited during routine AMM operations as users' exchange assets. Since assets held as open inventory have an additional utility of providing liquidity to the adopting AMM (DFMM), they are excluded from calculations seeking to quantify the efficiency of protocol's balance sheet. Guided by this notion, we introduce the concept of \lq\lq capital efficiency\rq\rq, defined as follows:
    
    \begin{definition}[Capital Efficiency]
        Capital efficiency ($\mathcal{E}_t$) is a measure quantifying the effectiveness of an asset's exclusive use to enable IDA protocol to achieve its dependability objectives (as stated in the introductory section), which ultimately affects the protocol's ability to continuously engage in work with an expected positive utility. It can be quantified as:
        \begin{equation}
            \mathcal{E}_t = \frac{CS^S_t}{\sum^n_{i=1} \mathcal{L}^{E^i}_t (\mathcal{C}^i_t) \cdot \frac{\sum_{i=1}^{n} \mathcal{L}^{E^i}_t (\mathcal{I}^i_t - \mathcal{I}^i_{{LP}_t})- \sum_{j=1}^{m} \mathcal{L}^{E^j}_t (\mathcal{I}^j_{LP_t} - \mathcal{I}^j_{t})}{\sum_{i=1}^{n} \mathcal{L}^{E^i}_t (\mathcal{I}^i_t - \mathcal{I}^i_{{LP}_t})}},
        \end{equation}    
    \end{definition}

    \noindent
    Additionally, we now define coverage ratio, a system parameter, as follows:
    \begin{definition}[Coverage Ratio]
        Coverage ratio ($R_t$) is the percentage of circulating IDAs liabilities secured by assets of the protocol.

        \begin{equation}
            R_t = \frac{A_t}{CS_t^S}
        \end{equation}   
    \end{definition}

    \noindent
    Finally, we recall the concept of utilisation rate, which we introduced in the motivating work\cite{abgaryan2023dynamic}.
    
    \begin{definition}[Utilisation Rate]
        The utilisation rate $U^i_t$ is a point-in-time asset-specific measure that represents the state of the liquidity pools, calculated as a ratio of open inventory position to the maximum open inventory position which the DFMM system can support.
        \begin{equation}\label{eq:util_rate}
            U^i_t = \frac{\mathcal{I}^i_t - \mathcal{I}^i_{{LP}_t}}{\frac{\mathcal{C}^i_{t}}{\varrho^i_t}},
        \end{equation}

        \noindent
        where $U_t^i$ represents the utilisation rate for pools with a positive open inventory position, which is equivalent to the $U_{{LHS}_t}^X$ state variable proposed in the motivating work; and $\varrho^i_t$ is the collateralisation rate for a paticular asset and signifies the proportion of unhedged open inventory that needs to be held as collateral in the vault to mitigate position-related inventory risk.    
    \end{definition}

\begin{figure}[H]
        \begin{center}
            \includegraphics[width=4.1in]{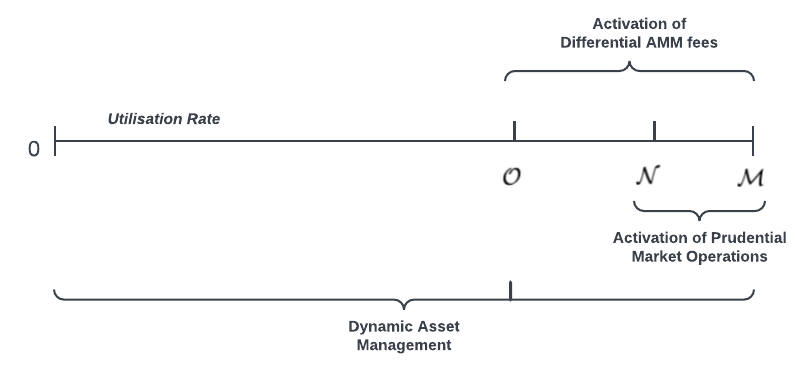}
            \caption{Activation of different levers.}
            \label{leverss}
        \end{center}
    \end{figure}

    \noindent
    Intuitively, the utilisation rate can be interpreted as an indicator of a liquidity pool's health\cite{abgaryan2023dynamic}, which the TALM policy can aim to manage using a pre-defined threshold.\\
    \\
    In essence, the system seeks to find a policy that enables us to: (i) maximise capital efficiency; (ii) minimise the probability that any of its constituent pool's are in a state of suboptimal utilisation rate; and (iii) extend fair accommodation to potential-hoarders to carry out corrective actions. Such dynamic policy ensures that, for all practical purposes, the \lq\lq balance sheet\rq\rq of the managed algorithmic asset backing its \lq\lq income statement\rq\rq, is not so static and fragile, that it is unable to promote IDA protocol's dependability when it's needed the most. Therefore, if the pre-programmed policy is executed as expected and liquidity pools are in a healthy state, i.e. within appropriate bounds explained in the forthcoming section, it can be said the protocol has efficiently allocated risk in a fully decentralised, dynamic and fairly incentivising manner. If however, the thresholds set are breached and any of the pools are in an unhealthy state, it indicates the system has reached an over-concentrated phase, and is at increased risk of destabilisation.\\
    \\
    In the subsections that follow, we describe how algorithmically programmed policies, actions of which are visualised in Fig.\ref{leverss}, use their differentiated levers to individually manage the asset and the liability side of the balance sheet to maximise the likelihood of the overarching ALM policy meeting its stated objectives.
    
    \noindent
    In Fig.\ref{leverss}, $\mathcal{O}$, $\mathcal{N}$ and $\mathcal{M}$ represent asset-specific utilisation rate-based policy thresholds, which determines which ALM tool would be used to assert IDA's TALM objectives.
    
\subsection{Dynamic Liability Management}
    In this section, we shift focus to TALM policy's dynamic liability management principles, which are affected by managing the utilisation rate of the pool ($U_t^i$), and aimed at achieving IDA's objectives, which includes dependability achieved through maintenance of healthy utilisation rates for all pools. To achieve this, two complementary levers, i.e. differential AMM fee (\lq\lq soft\rq \rq lever), and prudential market operations (PMO) (''hard" lever) are deployed using an intelligent lever selection policy.\\
    \\
    In eq.\ref{eq:util_rate}, we quantified the relationship between inventory, utilisation rate, LP holdings, and collateral. This can be rearranged to state the following: 
    \begin{equation}
        U^i_t \cdot \frac{\mathcal{C}_t^i}{\varrho_t^i} + \mathcal{I}^i_{{LP}_t} - \mathcal{I}^i_t = 0.
    \end{equation}

    \noindent
    Despite any variations in (realised) utilisation rate which may occur in any asset pool, the system targets a specific utilisation rate.

    \begin{definition} [Target Utilisation Rate]
        The target utilisation rate ($U^\ast_t$) represents the optimal rate that IDA and DFMM protocol are targeting, across all asset pools.     
  \end{definition}

    \noindent
    Note, that policy threshold ($\mathcal{O}_t$) introduced in Fig.\ref{leverss} can be thought of as a proxy of the target utilisation rate($ (U^{\ast}_t \approx \mathcal{O})$), representing a point on the Pareto optimal front containing the universe of solutions which optimise the multi-objective optimisation problem, comprised of balancing three objectives, i.e. maximising the stabilising time, maximising the capital efficiency, maximising the utility of IDA to an average agent, and maximising traded volume in the AMM.  

    \begin{equation}\label{eq:offchain}
        \text{Maximise}_{\mathcal{O}_t} f(\tau_t, \mathcal{E}_t, \mathcal{U}_{{\text{agent}_t}}, \mathcal{U}_{{{\text{Volume}_\text{AMM}}_t}})
    \end{equation}
    
    \noindent
    where $f(.)$ represents the value function. It is the solution to this optimisation problem, that is the (cross-asset)\footnote{Using a single target utilisation rate implies that we do not need to use asset-specific identifiers for variables reflecting utilisation rates. However, in forthcoming sections, we make the transition to quantifying asset-specific target utilisation rates using additional factors.} target utilisation.\\
    \\
    Naturally, if the current utilisation rate of a pool is different from the target utilisation rate, i.e. $U_t^i \neq U^{\ast}$, we can state that:
    \begin{equation}\label{eq:util_new}
        U^\ast_{t} \cdot \frac{\mathcal{C}_{t}^i}{\varrho_{t}^i} + \mathcal{I}^i_{{LP}_{t}} - \mathcal{I}^i_{t} \neq 0.
    \end{equation}

    \noindent
    Correspondingly the inventory level of the pool can be adjusted by $\Delta \mathcal{I}^i_t$ that facilitates the following equality:  
    \begin{equation}\label{eq:util_new2}
            U^\ast_{t} \cdot \frac{\mathcal{C}_{t}^i}{\varrho_{t}^i} + \mathcal{I}^i_{{LP}_{t}} - \mathcal{I}^i_{t} + \Delta \mathcal{I}^i_t= 0.
        \end{equation}

    \noindent
    On a ceterius paribus basis, to ensure that the system can achieve its target utilisation rate, the inventory level needs to adjusted by $\Delta \mathcal{I}^i_t$, which change can be calculated as follows:       

    \begin{equation}
           \Delta \mathcal{I}_t^i = \mathcal{I}^i_t -\mathcal{I}^i_{LP_t} - U^\ast_t \cdot  \frac{\mathcal{C}^i_t}{\rho^i_t}
       \end{equation}

    \noindent
    In the subsections that follow, we discuss IDA's soft lever, i.e. the differential AMM fee, the hard lever, i.e. the prudential market operations, and a dynamic liability lever selection mechanism, which are designed to enable pools to achieve their target utilisation rate.

\subsubsection{Differential AMM Fee}
    IDA is a dependable economic unit of measurement facilitating operations in multi-asset platforms, without hindering market forces which enable price discovery. For transactions specific to this asset, the protocol quantifies a special AMM fee ($\theta^i_{{S}_t}$), which seeks to catalyse its pools to achieve their target utilisation rate, which is the focal point of this section.\\
    \\ 
    This asset-specific fee is presented in the form of a piecewise function below:
    Now, the asset-specific differential fee is presented below in the form of a piecewise function:
    \begin{equation}\label{eq:difffee}
        \theta_{S_t}^i =
        \begin{cases}
            \theta_0 - (\theta_0 -\lfloor \theta \rfloor) \cdot \frac{U^i_t}{U^{\ast}_t}  & \forall U^{\ast^{+}}_{t} \geq U^{i^{+}}_t\\
             \lfloor \theta \rfloor \cdot (1 + D \cdot \frac{U_{t}^i}{U^{\ast}_{t}}), & \forall U^{\ast^{+}}_{t} < U^{i^{+}}_{t}\\
           \theta_{S_{t-1}} - (\theta_{S_{t-1}}- \lfloor \theta \rfloor) \cdot (1-\frac{U^i_t- U^{\ast}_t}{U^i_{t-1}- U^{\ast}_t})& \forall U^{\ast^{-}}_t < U^{i^{-}}_t,\\
            \lfloor \theta \rfloor + (\theta_0 - \lfloor \theta \rfloor) \cdot  (1-\frac{U^i_{t}}{U^{\ast}_{t}}) & \forall U^{\ast^{-}}_t \geq U^{i^{-}}_t,\\
        \end{cases}
    \end{equation}

    \noindent
    where superscripts \{+/-\} with the coverage ratio represent whether the trade is a purchase (or sale) of IDA from the trader's perspective; $U^{\ast}_t$ is the asset-specific target utilisation rate for pools with a positive open inventory position s.t. pools with a negative (or nil) open inventory position have $\theta_{S_t}^i = \theta_0$ where $\theta_0$ is the base fee (a system parameter); $ \lfloor \theta \rfloor$ is the minimum fee when the utilisation rate is at the threshold level for both purchase and sale of IDA (a system parameter); $D$ is a system parameter which quantifies the impact of deviation from the target utilisation rate on the fee. Finally note, that this fee can be computed in a continuous time setting, to quantify the fee incurred by a trader to execute a trade.
    
\subsubsection{Prudential Market Operations (PMO)}
    In this section, we discuss how IDA's TALM policy conducts prudential market operations to help stabilise the balance sheet.

    \begin{definition}[Hoarding]
        Hoarding is defined as the unproductive accrual of a finite asset.
    \end{definition}

    \noindent
    It must be specifically noted that unproductive accrual in the context of IDA protocol, refers to the action of acquiring, or continuing to hold, units of IDA when the system's pools are in an unhealthy state.

    \begin{definition}[Marginal Hoarder]
        A marginal hoarder is a hoarder whose overall utility derived from their accumulated assets is at such a low level that they are compelled to liquidate, or are logically expected to do so, once they receive information about the anticipated future cost of hoarding.
        \begin{equation}
            \mathcal{U}^H_{\text{hold}}(q) -  \mathbb{E}_{\text{future}}[\text{Hoarding}] \leq \mathcal{U}^H_{\text{sell}}(q),
        \end{equation}
        where $\mathcal{U}^H_{\text{hold}}(q)$ represents the net utility of the economically strained hoarder from their current asset holdings; $\mathbb{E}_{\text{future}}[\text{Hoarding}]$ is the anticipated future cost of hoarding; $\mathbb{E}_\text{future}$ is the expectation operator; and $\mathcal{U}^H_{\text{sell}}(q)$ is the utility the hoarder would derive from selling their assets.
    \end{definition}
    
    \noindent
    Naturally, agents of a digital asset ecosystem choose to hoard because of expectations of a favourable future price change (or have incentives that can be mapped onto as economic benefit), which may be at the cost of the finite asset's overall utility for the wider community; or in the absence of a favourable return expectation, simply an expectation of a future utility of an asset; or for \lq\lq known-unknown\rq\rq behavioural biases. In other words, if $\mathcal{U}(q)$ is the utility of the agent holding the economic resource, and $q<Q$ being the quantity suspected to be hoarded where $Q$ is the upper limit representing the total units of the finite asset available. We can state that the agent's utility $\mathcal{U}(q)$, where $\mathcal{U}^{'}(q)$ is strictly decreasing to reflect the diminishing marginal utility.\\ 
    \\   
    Hoarding is a source of risk that compromises the likelihood of IDA's success as a cross-chain transactional asset in the short-medium term, and in the long term, brings its very existence into question - because this impedes the adopting AMM's ability to absorb adverse movement in the market price of asset(s) backing its value.
    
    \begin{definition}[Prudential Market Operations (PMO)]
        Prudential market operations is the process of compelling conversion of a fraction ($H_t$) of the circulating supply of IDAs held by accounts owning the asset, into other digital assets, with the sole objective of restoring a pool's utilisation rate (in DFMM) to a sustainable level, i.e. if excessive hoarding has resulted in an unsustainable accumulation of risk.\\
        \\
        In other words, $PMO(.): nS \rightarrow \mathbf{a}$, where $\mathbf{a} = \{a_1, a_2, \dots, a_n\}$ represents a vector of allocation to the n-many (unique) digital assets available to convert IDAs to, based on pre-determined allocation rules, and $nS$ represents the number of units of the IDA.
    \end{definition}

    \noindent
    Execution of PMO(.) essentially removes users' discretion on which asset(s) would their IDA holdings be converted to, which can be seen as imposing a cost, as it may expose agents to unforeseen market risks from owning a potentially undesirable IDA asset, which has the effect of transforming the objective function of an agent to:
    \begin{equation}
        \mathbf{max}_q (\mathcal{U}(q) - \mathcal{H}(q)),
    \end{equation}

    \noindent
    where $q$ represents the number IDA assets held by an agent; $\mathcal{H}(q)$ represents the equivalent hoarding cost; and $\mathcal{U}(q)$ a general utility function from engaging in this penalised activity.\\
    \\
    Now, PMO triggers the conversion of a portion of IDA's circulating supply($CS^S_t$) to other assets, at predefined thresholds:
    \begin{definition}[PMO Thresholds]
        PMO thresholds (Fig.\ref{leverss})($\{\mathcal{M}_t, \mathcal{N}_t\}$) are defined using pool-specific utilisation rate,  where $\mathcal{M}$ is a system parameter indicating the level of utilisation rate when we compel the immediate conversion of IDA to an asset mix; $\mathcal{N}$ is the point of utilisation rate when the PMO policy is initiated (requiring periodic conversion) (see Fig. \ref{dynamiclevers}).\\
        \\
        At PMO's decision threshold $\mathcal{N}_t$, the system decides to deploy hard levers intended to compel action after $N_{h_t}$-many epochs \footnote{Epochs are defined in terms of fixed time intervals such as block time}, which was encouraged by soft levers of the dynamic liability management policy.
        \end{definition}

    \noindent
    Now, let us consider a function to capture the non-linear relationship between the utilisation rate, time spent in hoarding, and the implied financial impact of IDA's conversion to a basket of other assets.\\
    \\
    The imbalance of each pool of DFMM(with positive open inventory position), expressed by pool-specific utilisation rate ($U^i_t$) exceeding the threshold $\mathcal{N}_t$ can trigger the potential conversion for a portion of the circulating supply of IDA to the corresponding asset of the pool. We define this portion of IDA's circulating supply that is subject to conversion by $H^i_t$ (in \%), where $i$ indicates which pool(with positive open inventory) experiences an unsustainable utilisation rate and correspondingly to which asset IDA would be converted. Overall, the below-discussed mechanism is running for all pools independently and concurrently.\\
    \\
    $H^i_t$ can be computed as follows:
    
    \begin{equation}\label{Heqn}
    H^i_t = 
    \begin{cases}
        k \cdot (1+ {U^i_t}^{\min\left(\max\left(\frac{T}{T_0},1\right) - 1,1\right)}) \cdot T^{1 + \max\left(\frac{T}{T_0} - 1,0\right)} & \text{for } \mathcal{N}_t \leq U^i_t < \mathcal{M}_t,\\
        \min\left(\frac{-\mathcal{A}^i_t}{CS^S_t},1\right) & \text{for } U^i_t \geq \mathcal{M}_t.
    \end{cases}
    \end{equation}

    \noindent    
    where $T$ is the number of PMO rounds\footnote{PMO rounds are defined by $N_{h_t}$ many of epochs, where $N_{h_t}$ is the number of epoch that system waits to conduct converstion after PMO is triggered} during which the utilisation rate of a pool has remained above the threshold $\mathcal{N}_t$, $T_0$ is a system parameter representing the threshold number of PMO rounds after which the system increases the hoarding conversion rate with the increase of the utilisation rate and the number of PMO rounds during which the utilisation has remained above the threshold, $\mathcal{A}^i_t$ represents the accounting state of the open inventory positions in nominal terms (equivalent to $T_t^i$ variable described in the motivating work), which can be either positive or negative based on system's open inventory position. Note, that for the first interval, the conversion price is the market price, i.e. $P_{con} = P^{i^L}$ represents the price of i-th asset in the adopting AMM (DFMM), whereas for the second interval, the conversion price is $ P_{con} = \frac{\mathcal{I}^i_t - \mathcal{I}^i_{LP_t}}{-\mathcal{A}^i_t}$.\\
    \\
    Simply put, the first interval prioritises the predictability of conversion rate, time and price for the user, and in the second interval, prioritises IDA's system-wide dependability, by converting the system's liability to digital assets. In the first interval, if users rebalance the particular pool, decreasing the utilisation rate, there is no compelled conversion of assets. It's important to note that in the second interval, conversion relies on the available inventory in the liquidity pools. The conversion price is determined by dividing the open inventory position by the IDA-based accounting state of the pool (-$\mathcal{A}^i_t$). As a result, the conversion price may be equal to or lower than the market price, $P^{i^L} \geq \frac{\mathcal{I}^i_t - \mathcal{I}^i_{LP_t}}{-\mathcal{A}^i_t}$, creating the potential for losses for IDA holders, which is reasonable for holders that fail to act when the system was in the first interval state.\\
    \\
    Further, when the utilisation rate exceeds the hoarding conversion threshold $\mathcal{N}_t$ yet remains below $\mathcal{M}_t$ (the first interval of Eq.\ref{Heqn}), the system emits a signal delineating an impending conversion, subsequently waiting for $N_{h_t}$ epochs prior to advancing with the conversion at a designated conversion rate $H^i_t$. However, if following the emission of the signal, users promptly interact with the DFMM system to partially exchange their IDA for a different asset, and within $N_{h_t}$ epochs (following the initiation of the signal), enabling the system to efficaciously restoring the utilisation rate to a level below $\mathcal{N}_t$, the system rescinds the conversion announcement and refrains from executing any conversion subsequent to $N_{h_t}$ epochs. In essence, the system is not seeking to impose force conversion that it can on its users, rather, creates a sense of expectation of being converted, thereby encouraging holders (marginal hoarders) who have a marginally smaller utility of holding IDA, to convert to other asset to comply with practices which promotes long term dependability. If we are of the view that the utilisation rate imbaalance(hard lever) is a problem, but not such a severe problem that it should imply an exponentially increasing conversion rate($H^i_t$), and should taper off at a pre-defined maximum rate, which is an input for the system, then we can deploy the logistic function for the first interval.

    \begin{equation}
        H = \frac{k \cdot U_t^i}{1 + e^{-a \cdot (T - b)}}
    \end{equation}
    
    \noindent
    where  $a$ and $b$ are coefficients that configure the relationship between the number of PMO rounds when the pools were unhealthy and the conversion rate, and $k$ modulates the relationship between the utilisation rate of a pool and the conversion rate\footnote{In blockchain-based protocols, a decentralized governance mechanism can determine the function applied in each epoch, which determines the aggressiveness of the system depending on the emerging vulnerability.}.\\
    \\
    Now, upon the initiation of the conversion process, the system converts IDA into a variety of assets based on the healthiness of distinct DFMM pools at the prevailing market price, or if a pool's utilisation rate has exceeded the threshold $\mathcal{M}_t$, with a price which is sustainable (based on state of the inventory in the pool) \footnote{As soon as the conversion is triggered the system needs to decrease all the wallets holdings of IDA by $H^i_t$\%, and use those funds to convert into targetted assets. If the number of holding addresses is large, such direct distribution on the blockchain might suffer from scalability challenges, and related cost-inefficiencies. Therefore, to mitigate this, the implementation can consider the use of Merkle distribution\cite{merkeldist}, which is based on  Merkle Tree data structure, which optimises cost and scalability.}. After which, the system reallocates the newly converted assets from the open inventory to the LP positions, distributed to its holders in the form of LP tokens\footnote{LP tokens are digital assets that represent the holder's ownership in a particular liquidity pool.}. To help further exemplify this, given that IDA is converted into $\Delta \mathcal{I}^i_t$ units of asset $i$\footnote{Note, that $\Delta \mathcal{I}^i_t$ used in equations that follow, is distinct from its uses before, where it was previously used to indicate a change in inventory required to achieve a target utilisation rate.}, this subsequent operation culminates in the alteration of the open inventory position and the size of LP's inventory in i-th asset pool, in the adopting AMM (DFMM).

    \begin{equation}
        \text{OpenInventory} = \mathcal{I}^i_{t-1} - \mathcal{I}^i_{LP_{t-1}} - \Delta \mathcal{I}^i_t  \hspace{1cm}   \mathcal{I}^i_{LP_t} = \mathcal{I}^i_{LP_{t-1}} + \Delta \mathcal{I}^i_t.
    \end{equation}
    
    \noindent
    Finally, to overcome the impact of calculative agents attempting to game the liquidation auction process (elaborated upon in the motivating reference work), in-turn causing risk scenarios that IDA's TALM policy seeks to avoid, we can restrict the use of IDA during DFMM's liquidation auction for rebalancing.

    \subsubsection{Dynamic Liability Lever Modulation}
        In the preceding sections, we outlined key levers that constitute IDA's dynamic liability management policies. In this section, we discuss how we decide which of the factors enable us to dynamically select the most appropriate lever, i.e. whether velocity or treasury-based. These can be broadly categorised into two main buckets, which are as follows:
        
        \begin{figure}[H]
            \centering
            \includegraphics[width=4.1in]{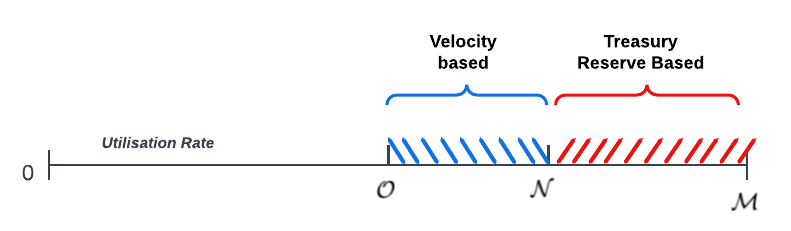}
            \caption{Dynamic lever selection.}
            \label{dynamiclevers}
        \end{figure}

    \noindent
    From Fig.\ref{dynamiclevers} it should be noted that the decision to select a particular lever is tantamount to modulating the distance of $\mathcal{N}_t$ from policy thresholds $\mathcal{O}_t$ and $\mathcal{M}_t$.\\
    \\
    Now, we consider two impact factors, one based on velocity, and the other based on the state of the treasury, to ascertain optimal $\mathcal{N}_t$ threshold.\\
    \\
    \textbf{a. Impact Factors: Velocity}\\
    \begin{definition} [Absolute velocity]
        Absolute velocity is defined as the rate at which a digital asset $i$ circulates within the DFMM  as a transactional asset, i.e.:
        \begin{equation}
            V^i_t = \frac{{TV^i_t}}{{CS^i_t}},
        \end{equation}
    \end{definition}
    
    \noindent
    where ${CS^i_t}$ represents the circulating supply of i-th asset, $V^i_t$ represents its absolute velocity, and $TV^i_t$ represents i-th asset's transactional volume in the adopting AMM (DFMM).\\
    \\
    In essence, IDA can be said to have a high velocity if it is used more frequently for trading, within DFMM, and a low velocity, if it is used less frequently. This distinction can be one of the factors in determining the optimal lever to deploy, as intended. However, if multiple digital assets are involved, one could employ the concept of relative velocity, as described below, to compare asset utilisation.    
        
    \begin{definition} [Relative velocity]
        Relative velocity refers to the rate at which a digital asset $i$ circulates within the DFMM as a transactional asset, on a normalised basis to other asset classes, i.e.
        \begin{equation}
            RV_i^t = \frac{V^i_t}{\overline{\mathbf{V}}},
        \end{equation}
    \end{definition}
    
    \noindent
    where ${\overline{\mathbf{V}}}$ is the average velocity, for n-many assets. In comparison to absolute velocity, relative velocity helps put the velocity of a particular asset in context to the velocity of others in the system. It may aid in future use cases involving contextual use of velocity, including in the pricing of derivatives listed on this asset.\\
    \\
    For IDA, we hypothesise that a correlated measure of changes in the velocity, is the variance of circulating supply($\psi(CS^S_t)$). This is because buying (or selling) of IDA leads to an increase (or decrease) of circulating supply, and thus, a high variance is not only an indicator of dominance as a means of executing trades, but also reflects the effectiveness of the transmission mechanism through which dynamic liability management can optimally deploy levers to achieve the target utilisation rate. So if IDA has high variance and therefore velocity, it can be an indication that participants are selling and buying a significant quantity of IDA (as a portion of its circulating supply). We hypothesise that rational IDA traders will choose pools with smaller fees, which helps progress individual pools towards their target utilisation rate. In other words, if IDA has a high variance, utilisation of 'soft' levers like differential AMM fees can be effective, whereas in other cases, 'hard' levers like prudential market operations like rules-based conversion, are required, as differential AMM fee can only have an impact when there are sufficient IDA based transactions.\\
    \\
    Recall that one of IDA's objectives (Sec.\ref{sec:intro}) is to promote harmonious behaviours, even amongst agents that can be financially seen to be adversarial to each other, and to maximise the time available to not only use IDA, but also to take corrective actions, if required. This naturally implies that IDA's objectives primarily prefer use of \lq\lq soft\rq\rq levers, i.e. differential AMM fees, before resorting to prudential market operations. Therefore, recall that Fig. \ref{dynamiclevers} included sequential trigger points $\mathcal{O}_t$ and $\mathcal{N}_t$, for soft and hard levers, respectively, whereby $\mathcal{O}_t \leq \mathcal{N}_t$. Along the same vein, we could interpret the difference between the two trigger points, i.e. $\delta_{{\mathcal{N} : \mathcal{O}}_t} = \mathcal{N}_t - \mathcal{O}_t$, to represent the expected effectiveness of soft levers to encourage corrective agent actions. Furthermore, as previously discussed, the variance of circulating supply can be seen as a proxy to IDA's velocity, and thus, imply how effective the differential AMM fees might eventually be, it is plausible to state that:
    \begin{equation}
        \delta_{{N:O}_t} \propto \psi^{z_t}(CS^S) \quad \forall z_t > 0,
    \end{equation}

    \noindent
    which can be extended to state a more detailed functional form, as follows:

    \begin{equation}
        \delta_{{\mathcal{N} : \mathcal{O}}_t}= min((\delta_{\mathcal{N}: \mathcal{O}}^{max} - \delta_{\mathcal{N}: \mathcal{O}}^{min}) \cdot l \cdot \psi_t(CS^S)^k + \delta_{\mathcal{N} : \mathcal{O}}^{min}, \delta_{\mathcal{N} : \mathcal{O}}^{max})
    \end{equation}
       
    \noindent
    where $\delta^{max}_{\mathcal{N} : \mathcal{O}}$ and $\delta^{min}_{\mathcal{N}:\mathcal{O}}$ are system parameters that signify the maximum and minimum permissible discrepancy between $\mathcal{O}_t$ and $\mathcal{N}_t$; and $l$ and $k$ are global parameters that govern the influence of the variance in circulating supply of IDA on $\delta_{{\mathcal{N} : \mathcal{O}}_t}$.\\
    \\
    \textbf{b. Impact Factor: Treasury State}\\
    \\
    In the event that the utilisation rate of pools remains undesirable after a pre-specified number of epochs, the adopting AMM's (DFMM) treasury resources are used to offer preferential pricing to attract rebalancing flows through the rebalancing premium auction. By extension, this naturally implies that the size of adopting AMM's (DFMM) treasury is an important indicator of its ability to support IDA. As such, threshold $\mathcal{N}_t$ is influenced by the treasury's state ($TR_t$).

    \begin{equation}
        \delta_{{\mathcal{M}:\mathcal{N}}_t} = max(\delta^{max}_{\mathcal{M}:\mathcal{N}} - (\delta^{max}_{\mathcal{M} : \mathcal{N}} - \delta^{min}_{\mathcal{M}:\mathcal{N}}) \cdot j \cdot TR_t^d, \delta^{min}_{\mathcal{M} : \mathcal{N}})
    \end{equation}

    \noindent
    where $\delta^{max}_{\mathcal{M}:\mathcal{N}}$ and $\delta^{min}_{\mathcal{M}:\mathcal{N}}$ are global parameters that signify the maximum and minimum permissible discrepancy between $\mathcal{M}$ and $\mathcal{N}$; and $j$ and $d$ are global parameters that govern the influence of the size of treasury on $\delta_{{\mathcal{M}:\mathcal{N}}_t}$.\\
    \\
    Note, that if $\delta_{{\mathcal{M}:\mathcal{N}}_t}$ and $\delta_{{\mathcal{N}:\mathcal{O}}_t}$, assume two different values for $\mathcal{N}_t$, we can calculate the threshold $\mathcal{N}_t$ as follows: 

    \begin{equation}
        \mathcal{N}_t = \beta \cdot (\mathcal{M}_t - \mathcal{O}_t - \delta_{{\mathcal{N} : \mathcal{O}}_t}) + (1-\beta) \cdot (\mathcal{M}_t -  \delta_{{\mathcal{M} : \mathcal{N}}_t})
    \end{equation}

    \noindent
    where $\beta$ is the weight of importance of the variance-based mechanism in the system to determine the optimal value for $\mathcal{N}_t$.

\subsection{Dynamic Asset Management}
    IDA's TALM principles seek to assure dependability, by maximising the number of epochs an agent has to convert their IDA assets to another asset, but also incentivise capital efficiency, as described in Sec.\ref{sec:talm}.\\
    \\
    Now, let us quantify the inventory level of an asset pool  ($\mathcal{I}^i_t$), using the size of collateral and utilisation rate:
    \begin{equation}
        \mathcal{I}^i_t = U^i_t \cdot \frac{\mathcal{C}^i_t}{\varrho^i_t} + \mathcal{I}^i_{{LP}_t},
    \end{equation}

    \noindent
    Similarly, the nominal value of the inventory level for asset $i$ can be calculated as follows: 
    
    \begin{equation}
        \mathcal{L}^{E^i}_{t}(\mathcal{I}^{i}_t)  = \mathcal{L}^{E^i}_{t} (U^{i}_t \cdot \frac{\mathcal{C}^i_t}{\varrho^i_t} + \mathcal{I}^i_{{LP}_t}).
    \end{equation}
    
    \noindent
    Specifically, note that IDA protocol continuously targets an optimal utilisation rate ($U^\ast_t = \mathcal{O}_t \geq 0$), using the adopting AMM (DFMM) and IDA-specific tools (incl. incentives), which are as follows:
    \begin{itemize}
        \item \textbf{Rebalancing auction}, of a reverse-Dutch kind, is triggered if the utilisation rate is above the target level to attract flows that reinstate the utilisation rate to the target level.
        \item \textbf{Margin liquidation process}, which is the second sequential step, is triggered when the rebalancing auction fails to reinstate the utilisation rate to a target level.    
        \item \textbf{Differential AMM fee}, the policy described in the previous section can be a potent tool in catalysing convergence of utilisation rate of individual pools to their target level.
    \end{itemize}

    \noindent
    We assume that with the introduction of IDA to the circulating supply, i.e. when the pools are at their target utilisation rate, we realise a transitorily-sticky open inventory position, such that any temporary perturbations are counterbalanced using the protocol's aforementioned levers, to enable convergence of the utilisation rate with its target\footnote{We make an implicit assumption that the aforementioned dynamics will realise asymptotically, provided there is satisfactory demand for the asset, and continuing viability of the system's levers to assert it's stated objectives.}.
    
    \begin{equation*}
        \sum_i \mathcal{L}^{E^i}_{t} (U^i_t \cdot \frac{\mathcal{C}^i_t}{\varrho^i_t}) + \epsilon^\prime \approx \sum_i \mathcal{L}^{E^i}_{t} (U^{\ast}_t \cdot \frac{\mathcal{C}^{i}_t}{\varrho^i_t}) + \eta^\prime \approx CS^S_t.
    \end{equation*}
    
    \noindent
    Now, we can compute the capital efficiency, as follows:
    
    \begin{equation}\label{eq:cet}
       \mathcal{E}_t \approx \frac{\sum^{i}_{i=1} \mathcal{L}^{E^i}_{t}(U^{i}_t \cdot \frac{\mathcal{C}^i_t}{\varrho^i_t})}{\sum^{i}_{i=1} \mathcal{L}^{E^i}(\mathcal{C}_{t}^i)} + \epsilon\approx \frac{\sum_i \mathcal{L}^{E^i}_{t}(U^{\ast}_t \cdot \frac{\mathcal{C}^i_t}{\varrho^i_t})}{\sum_i \mathcal{L}^{E^i}(\mathcal{C}_{t}^i)} + \eta,
    \end{equation}

    \noindent
    where $\epsilon$, $\eta$, $\epsilon^\prime$, and $\eta^\prime$ represent temporary and manageable perturbations. We posit that these temporary and manageable deviations result from transitory actions of market participants, which will eventually be compensated using one of the protocol's tools, and therefore, for ease of analytical convenience, these are omitted from equations that follow. In essence, target utilisation rate($U^{\ast}_t)$ can be used to achieve a target capital efficiency ($\mathcal{E}^{\ast}_t$).\\
    \\
    We now express IDA's expected persistent open inventory as a function of the size of collateral and target utilisation rate:
    
    \begin{equation}
        \mathop{\mathbb{E}}(I^i_t - I^i_{LP_t}) = f( C^i_t, U^{{\ast}}_t).
    \end{equation}

    \noindent
     Similarly, we can express assets backing IDA portfolio assets as weights, as follows:
     \begin{equation}\label{eq:assetwgts}
        w_t^i = \frac{A^i_t}{A_t}= f(\mathcal{\mathbf{C}}_t, \textbf{U}_t),
    \end{equation}

    \noindent
    where $\mathcal{\mathbf{C}}_t$ and $\mathbf{U}_t$ represent a vector of collateral levels in the long vault and target utilisation rates for all pools of the system, respectively; and $A^i_t$ is the nominal value of i-th asset in IDA asset base, which is assumed to be transitorily-sticky, and calculated as follows:
    \begin{equation}
        A^i_t=\mathcal{L}^{E^i}_t(I^i_t-I^i_{LP_t}) +\mathcal{L}^{E^i}_t(\mathcal{C}^i_t).
    \end{equation}

    \noindent
    Note, that if all assets are completely uncorrelated, and assuming that the collaterisation rate($\varrho^i_t$) and sLP behaviour-based risk assessment mechanism (discussed in forthcoming subsection) normalise the risk between different pools, the optimal utilisation rate for all pools could be the same ($U^\ast_t$), which is a system parameter, calculated off-chain by solving a multiobjective optimisation problem specified in Eq.\ref{eq:offchain}, which is restated below:

    \begin{equation*}
        \underset{\mathcal{O}_i \approx U^\ast_t}
        {\text{maximise}}
        f(\mathcal{E}_t, \tau_t, \mathcal{U}_{\text{agent}}, \mathcal{U}_{\text{Volume}_\text{AMM}}).
    \end{equation*}

    \noindent
    However, if we also consider the cross-asset correlation and accommodating new risk signals, pool-specific adjustment($\Delta U^i_t$) to the cross-pool target utilisation rate can be introduced, which leads to pool-specific optimal target utilisation rate ($U_t^{\ast^i} = U^\ast_t - \Delta U^i_t$). As such, we seek to minimise the portion of the IDA that would be converted in each timestep, which is equivalent to optimising the mix of assets backing IDA, as follows:
    
    \begin{equation}
    \begin{aligned}
        & \underset{\mathbf{C_t,\Delta U_t}}
        {\text{maximise}}
        && \frac{CS^S_t- \sum_i H_t^i \cdot CS_t^S}{CS^S_t}\\
        & \text{subject to}
        && ES_{1-\alpha} = -\frac{1}{{1 - \alpha}} \int_{-\infty}^{\text{VaR}_{1-\alpha}} x \cdot f(x) \, dx < \lceil ES_{1-\alpha} \rceil,\\
    \end{aligned}
    \end{equation}    

    \noindent
    where $ES_{1-\alpha}$ represents the expected shortfall at a pre-specified confidence interval (system parameter); $\lceil ES_{1-\alpha} \rceil$ is its threshold at the portfolio level; $\text{VaR}$ is the Value at Risk; $x$ is the portfolio loss and correspondingly $f(x)$ is the probability density function of portfolio's loss - quantifying the likelihood of different loss values; $\mathbf{\Delta U}$ is the vector of asset-specific adjustments of the on the $U^\ast_t$, which could be both positive and negative \footnote{We elaborate on the $\mathbf{\Delta U}_t$ in the upcoming sections}.
    
    \begin{definition}[Expected Shortfall (ES)]
        Expected shortfall ($ES_{1-\alpha}$), often referred to as conditional value at risk (cVaR) quantifies the average expected loss, if the losses stemming from a portfolio of financial assets breach the pre-set value at risk, at a given confidence interval $1-\alpha$.
        \begin{equation}
            ES_{1-\alpha} = \mathbb{E}(r_t | r_t > VaR(\alpha)).
        \end{equation}

    \noindent
    where $r_t$ is the realised loss. We chose this particular measure of risk because it specifically focuses on events that bring into question an asset's ongoing viability. In contrast, more popular measures like VaR, merely focus on the threshold of losses that one could experience in ordinary market conditions.
    \end{definition}
    
    \noindent
    In the optimisation problem stated above, we seek to maximise the portion of circulating supply that is not converted, using the decision variables $\mathcal{\mathbf{\Delta U_t}}$ and $\mathbf{C}_t$, which capture the pool specific utilisation rate adjustment and the size of collateral held in their respective margin vaults, respectively\footnote{An implicit assumption we make is that the vol-of-vol of asset prices are on par with the ability of the collateralisation rate to update, failing which, we would have a situation where market is moving too fast, well before IDA's TALM has an opportunity to react to evolving market conditions. Therefore, particular attention must be paid to the portfolio's variance $\sigma_p^2 = \sum_{i=1}^{n} w^{i^2}_t \sigma^{i^2}_t + \sum_{i=1}^{n} \sum_{j\neq i} w^i_t w^z_t \sigma^i_t \sigma^z_t \rho^{iz}$, where $\sigma^i_t$ is the standard deviation of returns of asset i, $\rho^{iz}$ is the correlation coefficient between returns of assets $i$ and $z$}.\\
    \\
    Dynamic asset management principles expressed in this section thus far, are implemented through a complementary pair of market and protocol-controlled levers, which are conditionally deployed using a pre-programmed and rules-based mechanism. While the market driven-lever facilitates the expression of the market view about individual asset risk, modulating collateral pool sizes($\mathbf{C}_t$), the protocol-driven lever enables incorporation of cross-asset correlation in the function which governs the optimal asset mix we are targeting, by modulating the asset-specific change in utilisation rate ($\mathbf{\Delta U}_t$).

    \subsubsection{Market-driven Lever}
\begin{figure}[H]
            \centering            
            \includegraphics[width=4.1in]{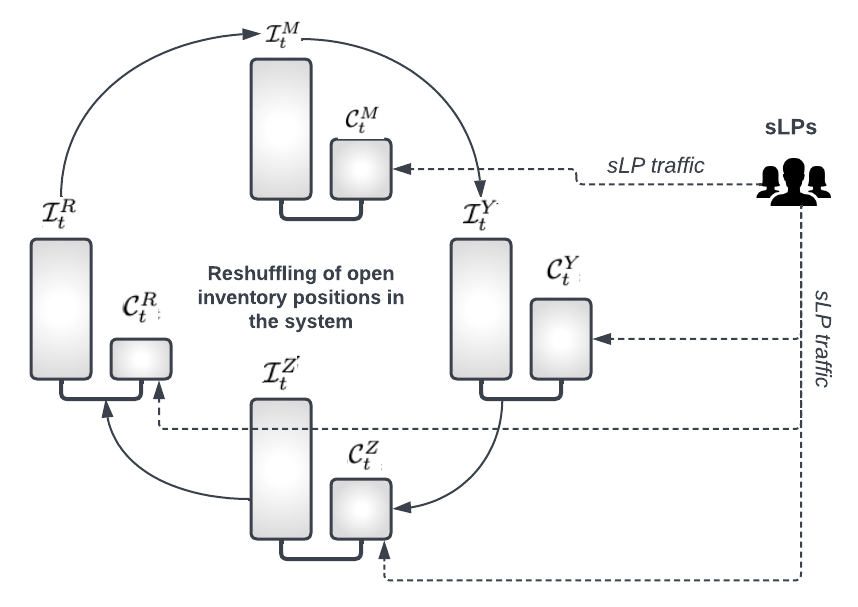}
            \caption{Schematics showing reshuffling of open inventory positions.}
            \label{shuffling}
        \end{figure}

        In this section, we seek to construct an optimal portfolio comprising the asset base of IDA protocol, with an emphasis on the expected shortfall of the individual asset, whilst the lever that follows in the next section singularly focuses on correlation.\\
        \\
        An informed quantification of expected shortfall requires learning the expected loss distribution of IDA's portfolio. Whilst realised prices from constituents of the asset pool would be informative, these would require augmentation with more nuanced risks being priced in the sLP market - which aides in rebalancing of the portfolio of assets backing IDA.\\
        \\
        Recall, that in the motivating work, we introduced the provision of risk-seeking agents that are native to the system, i.e. secondary liquidity providers (sLP). In essence, sLPs are agents who are incentivised to assess, price, and make markets in novel digital swaptions introduced in the same work, which enable (largely passive) primary LPs to obtain desired protection to their inventory risk. This means that at a high level, there is a class of agents, primary LPs, who are interested in maintaining limited exposure to any potential inventory risks resulting from market transactions, another class of agents - secondary LPs, who are interested in pricing the risk of a particular digital asset, which can be said to have a harmoniously adversarial nature.\\
        \\
        Further, being risk-seeking agents sLPs are adept at quantifying and pricing risks of any asset. And since we assume all market participants to be rational, it would not be an implausible assumption to state that sLPs will act upon any risk signal received, which will be transmitted by way of rebalancing their allocations to vaults which enables IDA to optimise its own delta exposures. The manner in which this \lq\lq rebalancing\rq\rq is effected, is by repricing the digital swaptions (by sLPs) with a range of statistical inputs to the model. In essence, sLPs can be thought of as mapping functions between the entire universe of relevant data emerging in concerned digital assets, and resulting risk-specific measures, to effect dynamic adjustment of collateral base.\\
        \\
        \textbf{sLP Market Dynamics}\\
        \\
        To further crystalise how dynamics developing in the sLP market can get nuanced risk information about IDA, let us delve into sLP market dynamics. Simply put, sLPs post bid and ask prices by depositing money in long and short vaults of digital assets, they wish to participate in providing limited risk transfer, using the bespoke digital swaption we have defined in Sec.\ref{sec:prelims}. In posting these bid ($P^{\partial i}_{b,t}$) and ask ($P^{\partial i}_{a,t}$) prices by submitting assets to short and long vaults (and changing the size of collateral ($\mathbf{C}_t$), they are essentially assessing the financial risk of the underlying digital asset they are trading, and contributing to the discovery of its derivative's fair value, uniquely represented with the superscript $\partial i$. Let this be represented as $FV^{\partial i}_t$, for an asset ${\partial i}$ at time $t$. Assuming frictionless markets, the sLP expects to be able to buy and sell at $FV^{\partial i}_t$, thereby, for ease of convenience, it can be stated that $|P^{\partial i}_{{b,t}} - FV^{\partial i}_t|$ represents the theoretical profit. To maintain competitiveness with other sLPs in the market, they have to constantly update their view of the market by updating their beliefs of factors that enable fair value calculations of the digital swaption contract. Now, if we assume, that a Black-Scholes model is repurposed to fit the purpose, then, sLPs would be required to ascertain - their expectation of the underlying digital asset's price at the point of inception of the aforementioned contract, during the life of the contract, and especially, at maturity. This is in addition to having to quantify the implied volatility inputs, the yield (cost) associated with simply holding the asset, slippage or transaction cost, the size of open inventory for any given asset and its impact on the price of the digital swaption, among other higher-order statistics. Whilst these measures are used to update the size of deposits posted in the long or short vault of a particular digital asset, IDA protocol can infer the implied changes in several variables by observing how sLPs are adjusting their exposure to any given underlying. This correlative information is in addition to causal factors which can be learned from how the pLPs are reacting to changes in the sLP market, which will naturally compel a rebalancing of IDA's asset mix in the direction of an optimal mix.\\
        \\
        In Fig. \ref{shuffling} that follows, we demonstrate how sLP behaviour which leads to changes in the collateral pool's size, can facilitate the reshuffling, and diversification of assets backing the IDA. 

    \subsubsection{Protocol-driven Lever}
        The sLP-based mechanism elaborated in the previous section enables optimisation of the asset mix backing the portfolio, without considering the cross-asset correlations. In this section, we complement a lever driven by the market's natural force - which act as a natural lever to assert IDA's dynamic asset management principles, with a protocol-driven lever, which works based on risks gleaned from cross-asset correlation.\\
        \\
        Specifically, this approach adopts a two-pronged approach, the primary of which is to use a correlation matrix derived using smart contract-based calculations (or inputs from an oracle), to enforce an upper-bound on utilisation rate for each of the assets in IDA. Backing this up would be a decentralised governance-based mechanism, to tackle spurious inputs which may be hard to foresee ex-ante.\\
        \\
        Whilst the latter assures the availability of customisation of the response function, the former enables us to apply a dynamic measure to manage risks emerging from the composition of assets backing IDA. If all assets were perfectly uncorrelated to each other, we could apply a uniform target utilisation rate to them($U^{\ast}_t$), however since there isn't a perfect lack of correlation on a collateral-adjusted basis, we have to adjust the target utilisation rates for each asset pool, that are acceptable to IDA protocol.\\
        \\
        Furthermore, at a portfolio level correlation tends to have the affect of amplifying volatility, which means, in it's truest sense, the collective asset mix could be more volatile than the sum of its parts. To quantify this, we state the formula for portfolio volatility: 
        \begin{equation}
            \sigma_p = \sqrt{w_1^2 \sigma_1^2 + w_2^2 \sigma_2^2 + 2w_1w_2\rho\sigma_1\sigma_2}.
        \end{equation}

        \noindent
        In the portfolio-level volatility quantified above, it can be observed that as long as $\rho > 0$, assets are correlated to each other, leading to higher portfolio-level volatility.\\
        \\
        As a starting point, we quantify the weighted correlation of a particular asset with all the other assets in the IDA mix, as follows:
        \begin{equation}
            \rho^i_t = \frac{\sum_z \rho^{i,z} \cdot \mathcal{C}^z_t}{\sum_z \mathcal{C}^z_t}
        \end{equation}
        
        \noindent
        which gives us the collateral-weighted correlation of the i-th asset, using a pairwise correlation of this asset, with all of the other assets in IDA asset base. Now, using this correlation, we can quantify the asset-specific utilisation rate adjustment ($\Delta U^{i}_t$), which helps adjust the utilisation rate for correlation, when assessing the utilisation rate, as follows:

        \begin{equation}
            \Delta U^{i}_t =
            \begin{cases}
               \alpha_{-}(w^i_t) \cdot \tan\left(\beta (\rho^i_t)^\gamma\right)^\phi & \text{for } \rho^i_t < 0 \\
               \alpha_{+}(w^i_t) \cdot \tan\left(\beta (\rho^i_t)^\gamma\right)^\phi & \text{for } \rho^i_t > 0,\\
            \end{cases}
        \end{equation}

        \noindent
        where $\Delta U^{i}_t$ is the adjustment as defined above, $\rho_t^i$ is the time-dependent correlation of a particular asset in the IDA mix, and $\alpha, \beta, \gamma$ are responsiveness coefficients which are determined using the response of the market to IDA deployment of levers. Specifically, $\alpha(w^i_t)$ is the scaling factor which enables us to differentiate the response function between positive and negatively correlated assets and is determined also using the impact of open inventory; $\beta$ stretches or compresses the impact of changes in correlation on the output, and thus a larger value of $\beta$ leads to more rapid oscillations; $\gamma$ is the exponent which controls the steepness of the function, s.t. a If $\gamma < 1$, it amplifies small values of $\rho^i_t$ and reduces the impact of larger values, resulting in a steeper curve. Conversely, if $\gamma > 1$, it reduces the effect of small values and amplifies larger values, leading to a flatter curve; $\phi$ is the scaling coefficient which controls the concavity or convexity of the output. We believe such piecewise and parametric functional form enables us to retain maximum flexibility, whilst having a well-defined mechanism to respond to changes in observed correlation. However, we will simulate, and update, the form if need be.\\
        \\
        In the forthcoming revisions, we derive admissible ranges of all parameters for differing market circumstances.

    \subsubsection{Dynamic Asset Lever Modulation}
        Exploring the interval function, this section delves into how cVaR-based (market-driven) and correlation-based (protocol-driven) levers are utilised in managing IDA's portfolio.\\
        \\
        In a turbulent market environment, defined as a situation where assets comprising IDA exhibit high volatility and a higher likelihood of extreme adverse events, the higher moments of the returns distribution capture key risks. This focus on tail risk is captured using expected shortfall, as in scenarios with a high $cVaR_\alpha$, we can apply a higher scaling factor. Contrastingly, in a stable market environment, characterised by periods where IDA assets experience low volatility, leading to a lower likelihood of extreme negative events, the scaling factor could be low.

\section{Conclusion and Future Work}
\begin{figure}[H]\label{fig:idaclusters}
        \centering
        \includegraphics[width=4.1 in]{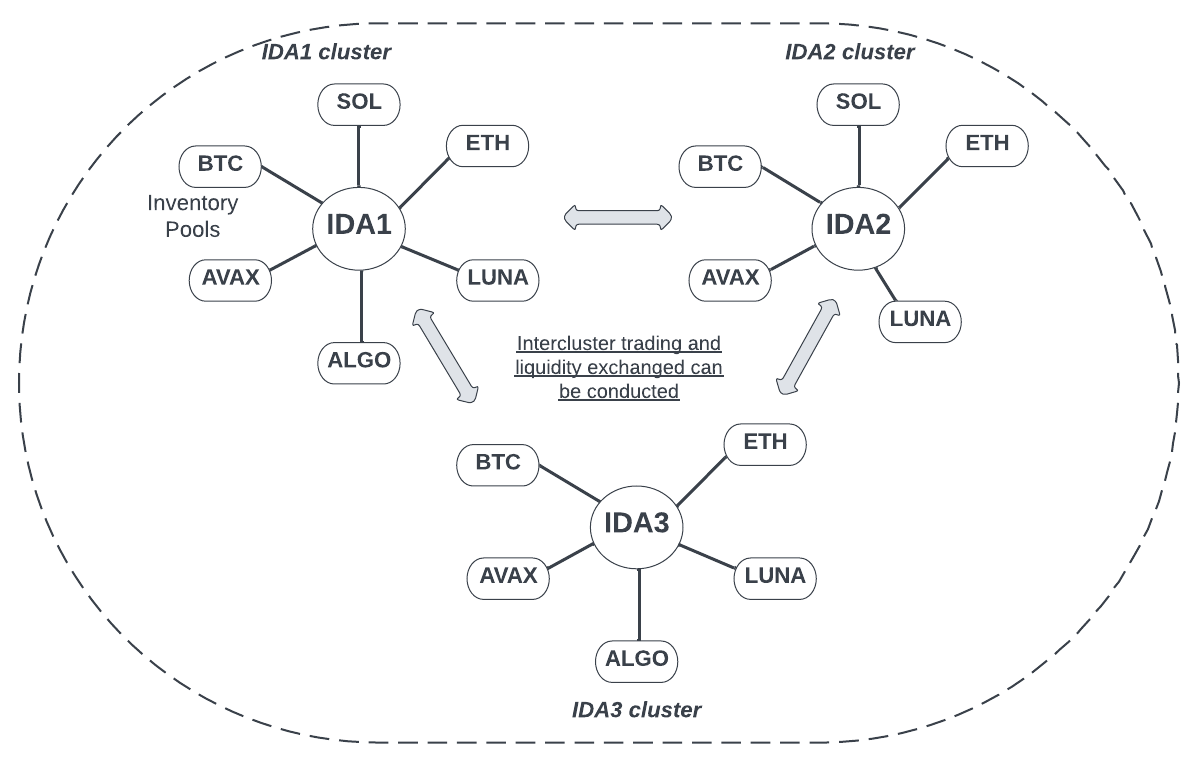}
        \caption{IDA Clusters}
    \end{figure}

    In this work, we have introduced an innovative concept known as an Intermediating Decentralised Asset (IDA). IDA is managed through a meticulously defined rules-based mechanism, serving as an intermediary asset that doesn't impede market forces, taking into account the value of assets such as indices, currencies, other assets, or even a bundle of assets used as anchors for IDA's value. The dynamic asset and liability management policy associated with IDA enables its holders to seamlessly convert their holdings into other assets. Separately, this proposition also lays the foundation for integrating the protocol with a data oracle service, enabling the development of a community-driven suite of tradeable indices.\\
    \\
    In forthcoming revisions, we will provide the results of our simulation, along with formal proofs, and introduce how communities across different blockchain ecosystems can create new instances of IDA protocol, by inviting pLPs, sLPs, and traders, to participate. This would enable them to use different assets as the peg driving their own implementations of IDA, which we refer to as IDA clusters (Fig.\ref{fig:idaclusters}). We will also study the dynamics between multiple pegs, the role of relative velocity in prudential market operations, and general policy-making for inter-cluster harmony.
    
\bibliography{main.bib}
\bibliographystyle{plain}

\end{document}